%% file: ifacconf.tex
\documentclass{ifacconf}

\usepackage{graphicx}      
\usepackage{natbib}        
\usepackage{subcaption}
\usepackage{algorithm}
\usepackage{algpseudocode}
\usepackage{amsmath}
\usepackage{url}            
\usepackage{placeins}

\begin{document}
\begin{frontmatter}

\title{Fairness Evolution in Continual Learning for Medical Imaging} 

\author[First]{Marina Ceccon} 
\author[First]{Davide Dalle Pezze} 
\author[Third]{Alessandro Fabris}
\author[First]{Gian Antonio Susto}

\address[First]{Università degli studi di Padova, 
   Padova, PD 35137 IT (e-mail: marina.ceccon@phd.unipd.it, davide.dallepezze@unipd.it, gianantonio.susto@unipd.it).}
\address[Third]{Università degli studi di Trieste, 
   Trieste, TS 34127 IT (e-mail: alessandro.fabris@units.it)}

\begin{abstract}                
Deep Learning has advanced significantly in medical applications, aiding disease diagnosis in Chest X-ray images. However, expanding model capabilities with new data remains a challenge, which Continual Learning (CL) aims to address.
Previous studies have evaluated CL strategies based on classification performance; however, in sensitive domains such as healthcare, it is crucial to assess performance across socially salient groups to detect potential biases.
This study examines how bias evolves across tasks using domain-specific fairness metrics and how different CL strategies impact this evolution.
Our results show that Learning without Forgetting and Pseudo-Label achieve optimal classification performance, but Pseudo-Label is less biased.
\end{abstract}


\end{frontmatter}
\section{Introduction}

In recent years, Deep Learning (DL) models have been successfully applied to various domains in the medical field, including pathology classification, anatomical segmentation, lesion delineation, image reconstruction, synthesis, registration, and super-resolution \citep{umirzakova2023medical}, exhibiting impressive performance across these tasks \citep{celard2023survey}.

Despite these advancements, DL models encounter significant challenges when trained on real-world data, especially in dynamic domains such as medical imaging. In these settings, continual updates in data distribution—due to emerging diseases, new imaging techniques, or shifting patient demographics—can result in substantial distributional shifts \citep{kumari2023continual}. Adapting to such changes is critical for model reliability and clinical relevance. However, fine-tuning on new data leads to catastrophic forgetting, where prior knowledge is overwritten \citep{kirkpatrick2017overcoming}. Conversely, retraining models from scratch is often infeasible due to high computational costs and privacy concerns related to storing or accessing old patient data \citep{dalle2023multi}.

Continual Learning (CL) has emerged as a promising solution to this challenge, offering a framework that enables models to adapt to evolving data streams while preserving prior knowledge. Past studies have explored CL strategies in medical imaging, mainly focusing on optimizing classification accuracy \citep{akundi2022incremental, lenga2020continual, singh2023class}. However, in the context of sensitive medical data, accuracy alone is insufficient. It is equally important to assess model fairness, as DL systems may exhibit performance disparities across demographic groups defined by protected attributes such as age, ethnicity, gender, and socioeconomic status \citep{seyyedkalantari2020chexclusionfairnessgapsdeep}. These disparities can lead to unequal care or misdiagnosis for vulnerable populations, highlighting the need to incorporate fairness into CL evaluation.

In this study, we analyze the evolution of bias across successive tasks using fairness metrics and investigate how different CL strategies influence bias progression over time. Specifically, we consider a class-incremental learning scenario using two widely recognized chest X-ray classification datasets: CheXpert (CXP) \citep{irvin2019CheXpert} and ChestX-ray14 (NIH) \citep{Wang_2017}. For both datasets, we construct a stream of five tasks, each involving two or three pathologies, covering 12 total pathologies in CXP and 14 in NIH. This setup allows us to study both classification performance and fairness trends as new diseases are gradually introduced.

Our contributions can be summarized as follows: \begin{itemize} \item We introduce the analysis of fairness metrics in a CL setting for medical imaging. \item We examine the evolution of bias throughout the task stream using the widely adopted CXP and NIH datasets in a class-incremental learning scenario. \item We compare the impact of different CL strategies on fairness metrics, highlighting their varying effects on bias mitigation. \end{itemize}

Our paper is structured as follows. In Sec. \ref{sec:Related_Work}, we review the existing literature on CL, algorithmic fairness, and their intersections within the medical domain. Sec. \ref{sec:ExperimentalSetting} details the considered scenario, along with the metrics and methodologies employed. In Sec. \ref{sec:results}, we present and analyze the experimental results. Finally, in Sec. \ref{sec:ConclusionFutureWork}, we discuss our findings and outline potential directions for future research.

\begin{figure*}[thbp] \centering \includegraphics[width=0.75\linewidth, trim = 0 0 0 0]{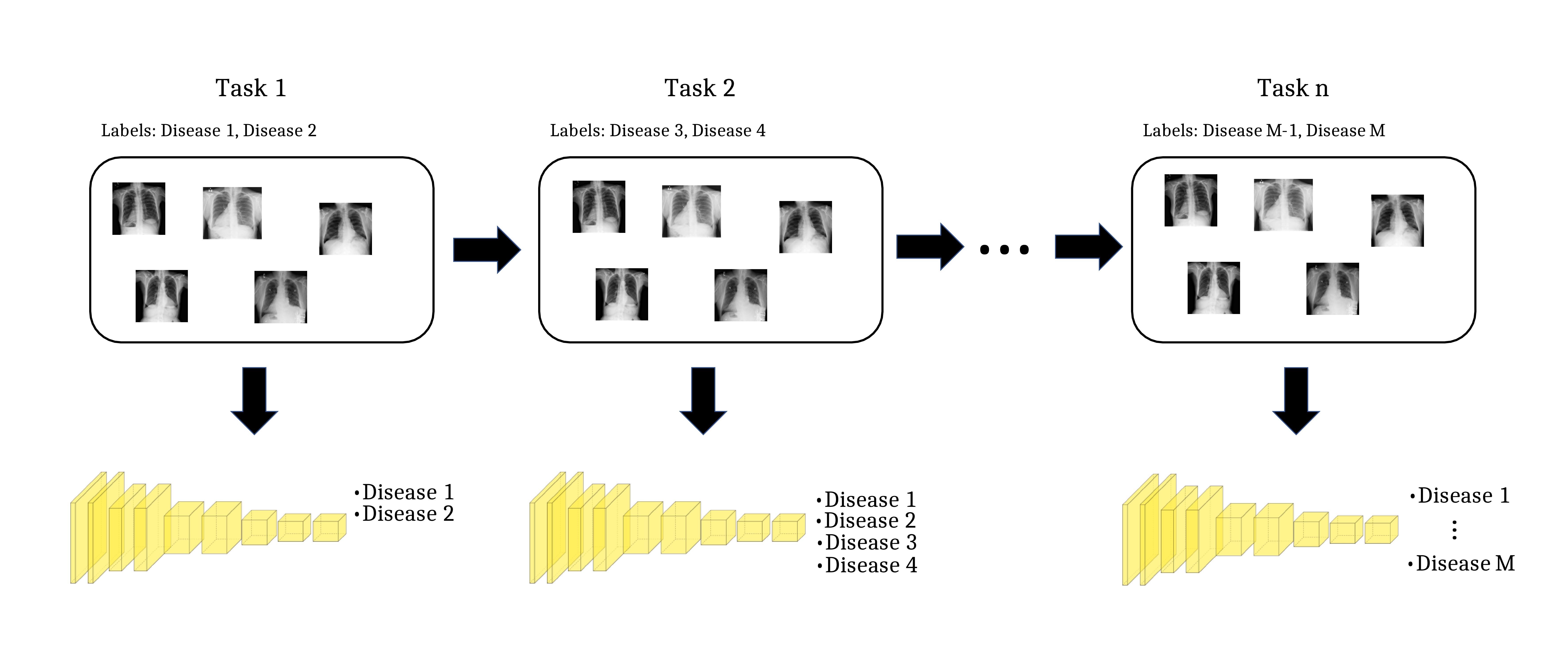} \caption{An example of the Continual Learning setting studied to evaluate fairness in the medical domain. In this setting, the model needs to adapt to the evolving medical knowledge by incorporating newly labeled diseases that appear over time.} \label{fig:CL_scenario} \end{figure*}

\section{Related Works}
\label{sec:Related_Work}

\subsection{Fairness in the Medical Domain}
\label{subsec:Fairness}

Machine Learning and Deep Learning models used in real-world decision-making may exhibit bias when handling sensitive attributes \citep{FairnessBook}, potentially leading to discriminatory outcomes for minority groups. To tackle this, fairness has emerged as a field in Artificial Intelligence focused on identifying and mitigating bias to develop fairer models.

Related to Fairness in the medical domain, \cite{seyyedkalantari2020chexclusionfairnessgapsdeep} analyze biases in pathology classifiers trained on chest x-ray datasets, evaluating performance across sex, age, race, and insurance type. Their findings show systematic disadvantages for females, Hispanic patients, Medicaid recipients, and younger patients. Similarly, \cite{zhang2022improving} train binary classifiers on MIMIC-CXR and CheXpert to predict the conditions Pneumothorax and Fracture. Their main finding is that while fairness-driven methods improve group fairness, they do so at the cost of reduced performance for all groups. Finally, \cite{weng2023sexbasedphysiologicaldifferencescause} investigate bias in deep learning models, hypothesizing that breast tissue causes underexposed lung regions and thus reduces model performance. By limiting training to one image per patient, they improve fairness without significantly harming accuracy.

\subsection{Continual Learning in the Medical Domain}
\label{subsec:introCL}

In conventional machine learning, models are trained on static datasets, which can lead to performance degradation when encountering novel data. Continual Learning (CL) addresses this by enabling models to incrementally learn a stream of tasks, though it introduces the challenge of catastrophic forgetting—where performance on earlier tasks deteriorates \citep{lesort2019continual}.

In CL, models learn from a sequence of tasks without forgetting prior knowledge, addressing the limitations of static training. CL is typically categorized into Domain Incremental Learning (DIL), where the input distribution shifts but class labels remain the same; Class Incremental Learning (CIL), where new classes appear without task identifiers; and Task Incremental Learning (TIL), where task identities are known \citep{lesort2019continual}.

Common CL strategies include rehearsal-based methods, which retain samples from past tasks (e.g., Experience Replay \citep{rolnick2019experience}); regularization-based methods, which constrain updates to preserve past knowledge (e.g., Learning without Forgetting \citep{li2017learning}); and architecture-based approaches, which dynamically modify network structure \citep{rusu2016progressive}.

In the medical domain, machine learning models must often adapt to new knowledge while preserving prior information. Changes in the environment or medical equipment can introduce distribution shifts in input data, affecting model performance \citep{lenga2020continual}. Moreover, new diseases may emerge or be retrospectively labeled after initial training \citep{singh2023class}. To address these challenges, research has explored continual learning (CL) applications in medical settings. \cite{singh2023class} introduce three tasks, in a CIL scenario, covering 12 classes using replay, and \cite{akundi2022incremental} propose a distillation-based method across five sequential tasks. \cite{10943697} further explore a New Instances and New Classes scenario, combining distillation and rehearsal.

\subsection{Fairness in Continual Learning} \label{subsec:FairnessCL}

Recent research has increasingly addressed fairness within continual learning settings. \cite{truong2025falconfairnesslearningcontrastive} propose FALCON, a method that employs contrastive clustering and attention mechanisms to mitigate bias during semantic scene segmentation. \cite{chowdhury2023sustaining} develop FaIRL, which sustains fairness across sequential tasks by controlling representation compression. Similarly, \cite{Churamani_2023} apply domain-incremental continual learning to facial expression recognition, employing continual adaptation for bias mitigation.

Despite these advances, to the best of our knowledge, no prior work has systematically evaluated or compared the fairness performance of continual learning methods on clinical data. This study addresses this gap by benchmarking multiple continual learning algorithms on chest X-ray classification tasks, assessing both predictive accuracy and fairness across demographic subgroups.

\section{Experimental Setting} \label{sec:ExperimentalSetting}

\subsection{Considered scenario} \label{subsec:our_scenario}

We model a medical imaging scenario in which a computer-aided diagnosis system assists specialists in interpreting X-ray scans. The system is continually updated to accommodate an expanding set of pathologies, with developers adding newly annotated images and organizing them into tasks for sequential improvement.

We consider a Class-Incremental Learning (CIL) setup using the CXP dataset \citep{irvin2019CheXpert} and the NIH dataset \citep{Wang_2017}. As in typical multi-label continual learning \citep{dalle2023multi}, information about previously learned classes is omitted from new tasks, even if they still appear in the images. This mirrors challenges in Object Detection and Semantic Segmentation within Continual Learning 
\citep{cermelli2020modelingbackgroundincrementallearning}.

For both datasets, we define a stream of 5 tasks, each linked to 2 or 3 pathologies. Following prior work in continual object detection \citep{shmelkov2017incremental}, each task includes only images with at least one relevant pathology. Tasks may contain overlapping images, depending on pathology correlation. We exclude “No Finding” images, as they are not associated with any pathology. Only one image per patient is included, following evidence that this improves model fairness without substantially harming classification performance \citep{weng2023sexbasedphysiologicaldifferencescause}.

\subsection{Evaluated methods} We consider several Continual Learning (CL) strategies:

\begin{itemize}
    \item \textbf{Fine-Tuning}: Sequential training on new data without mechanisms for retaining prior knowledge. Typically regarded as the lower CL performance bound.
    \item \textbf{Replay} \citep{rolnick2019experience}: We use a 50\% mix ratio and a memory buffer size of 3\% of the original dataset. Samples are stored and replayed uniformly at random.

    \item \textbf{LwF} \citep{li2017learning}: Uses the combined loss $L = L_1 + \tau L_2$, where $L_1$ handles the current task and $L_2$ is the distillation loss. We set $\tau=2$ as in \cite{li2017learning}.

    \item \textbf{Pseudo-Label} \citep{guan2018learn}: For each class, we determine a threshold that maximizes the F1 score on the validation set of its corresponding origin task. The teacher model’s outputs for previously learned classes are then binarized using these class-specific thresholds.

    \item \textbf{LwF Replay}: Combines Replay with LwF, using the same hyperparameters and sampling strategy as the individual methods.

    \item \textbf{Joint Training}: Trains the model on all tasks simultaneously, assuming access to the full dataset at once. While not a CL method, it serves as an upper-bound baseline unaffected by catastrophic forgetting.

\end{itemize}

\subsection{Evaluation metrics} \label{subsec:evaluation_metrics} To assess performance, we use ROC AUC, a standard metric for classification tasks. It is computed by plotting the True Positive Rate (TPR) against the False Positive Rate (FPR) across various thresholds. Given the task stream setting, we report the average AUC over all pathologies from all tasks seen up to a given point.

For fairness, we use the Equality of Opportunity (EO) metric, which evaluates TPR disparities across demographic groups. This addresses the problem of underdiagnosis in minority populations \citep{seyyed2021underdiagnosis}, where models often produce lower TPRs for disadvantaged groups. EO for pathology $i$ (in task $j$) is defined in Eq.~(\ref{eq:eo}), with $\sigma_j$ as the task’s test set, $\hat{y}_i$ as the model’s prediction, $y_i$ the ground truth, $a$ the advantaged group, and $d$ the disadvantaged group:

\begin{align}
    \text{EO}_{i} &= {\Pr}_{\sigma_j}(\hat{y}_{i}=\oplus \mid s=a, y_i = \oplus) \notag \\
    &\quad - {\Pr}_{\sigma_j}(\hat{y}_{i}=\oplus \mid s=d, y_i = \oplus). \label{eq:eo}
\end{align}

Since it measures a difference in TPRs, we additionally refer to it as \textit{TPR gap}. We examine fairness across gender and age. Specifically, we compare performance between males and females, and across age groups: 0–20, 20–40, 40–60, and 60–80. Males are treated as the advantaged group; for age, patients under 20 are advantaged, while those over 60 are disadvantaged. As with AUC, we compute EO over all tasks up to $j$ and report the average.

\section{Results}
\label{sec:results}
Here, we present the experimental results obtained on the CXP dataset. The corresponding outcomes for the NIH dataset are summarized in Table \ref{tab:results}. While the analysis focuses on the CXP dataset, the findings generalize to NIH due to the consistency observed across both datasets.

\subsection{Analysis on the classification performance}
\label{subsec:results_roc}
\begin{figure}[thbp]
  \centering
    \centering
    \includegraphics[width=0.35\textwidth]{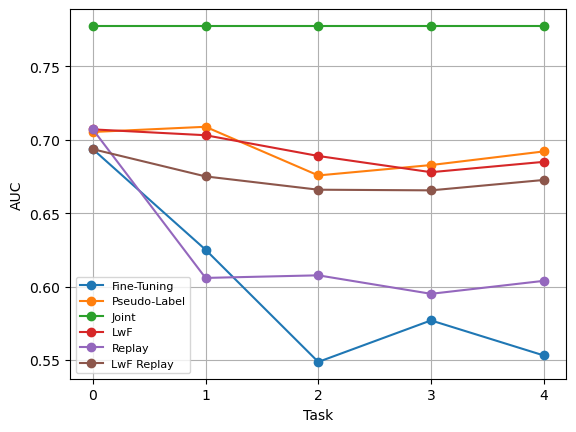}
    \caption{AUC metric, evaluated on each strategy, averaged on all the pathologies seen so far (CXP).}
    \label{fig:AUC_ROC_CXP}
  \label{fig:AUC_CXP}
\end{figure}

As shown in Fig. \ref{fig:AUC_ROC_CXP}, the model trained with joint training on the CXP dataset achieves an average AUC of 0.78, comparable to the state-of-the-art \citep{seyyedkalantari2020chexclusionfairnessgapsdeep}. The slight drop may stem from excluding “No Finding” images and limiting to one image per patient.

Fine-Tuning fails to preserve previously learned knowledge: adapting to new tasks degrades the AUC for earlier pathologies, reducing the overall average. Similarly, Replay struggles in this class-incremental multi-label setting due to interference—also observed in incremental object detection \citep{shmelkov2017incremental}—as identical images may appear in different tasks with conflicting labels.

In contrast, LwF and Pseudo-Label mitigate forgetting, helping the model retain earlier classes while learning new ones. They achieve average AUCs of 0.68 and 0.69. Despite improving over Replay, a gap remains between Pseudo-Label and the upper bound set by joint training.

Lastly, LwF Replay yields suboptimal performance, slightly below both LwF and Pseudo-Label. This degradation is attributed to interference introduced by replayed samples in the multi-label setting: while some samples reinforce learning, others conflict with the current task data.

\subsection{Analysis of the fairness evolution on the gender attribute}
\label{subsec:results_fairness_gender}

In this paragraph, we conduct disaggregated analyses of CL methods across the gender attribute to identify potential fairness disparities.

\begin{figure}[htbp]
  \centering
  \begin{subfigure}[b]{0.35\textwidth}
    \includegraphics[width=\textwidth]{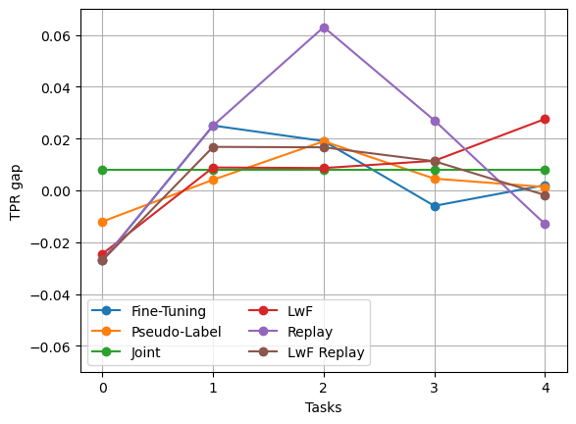}
    \caption{Gender EO on CXP for all the considered CL strategies.}
    \label{fig:TPR_CXP}
  \end{subfigure}
  \hfill
  \begin{subfigure}[b]{0.35\textwidth}
    \includegraphics[width=\textwidth]{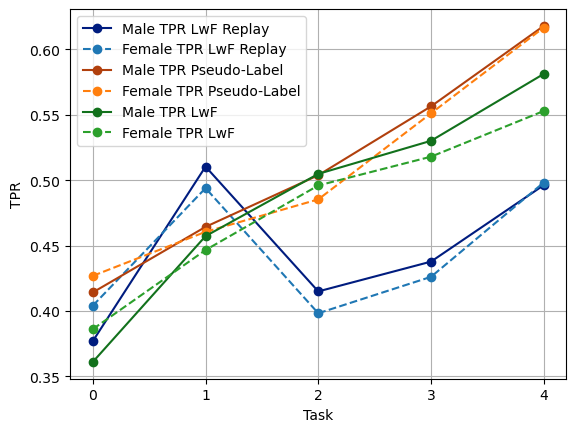}
    \caption{Male and female TPR evolution over the task stream of the three best CL strategies.}
    \label{fig:sex_TPR-CXP}
  \end{subfigure}
  \caption{Fairness metric results on CXP.}
  \label{fig:summaryTPR}
\end{figure}

It is important to note that, among the CL methods examined, we focus our analysis of fairness metrics on those demonstrating satisfactory AUC performance—specifically, LwF, Pseudo-Label, and LwF Replay. This focus is justified by the principle that fairness evaluation is meaningful only when the model maintains sufficient accuracy and mitigates catastrophic forgetting.

In the case of joint training on the entire dataset, previous studies have shown that models trained on CXP and NIH exhibit bias favoring male patients \citep{seyyedkalantari2020chexclusionfairnessgapsdeep}. In our setting, although the average TPR is still higher for males, the observed gap is smaller—only 0.008. This discrepancy from prior work may result from limiting the dataset to one image per patient, a strategy shown to reduce performance disparities \citep{weng2023sexbasedphysiologicaldifferencescause}, as well as from excluding “No Finding” images.

While the gap is minimal in this static setting, it remains essential to assess whether this trend holds in the Continual Learning scenario. Fig. \ref{fig:TPR_CXP} shows the EO between male and female patients across all methods, while Fig. \ref{fig:sex_TPR-CXP} presents the TPRs for male and female patients for LwF, Pseudo-Label, and LwF Replay across all tasks. For LwF, from the second task, male TPRs are consistently higher, resulting in a stronger EO than observed in joint training, potentially indicating underdiagnosis of women. In contrast, for Pseudo-Label, the EO fluctuates across tasks but converges toward zero. Similarly, LwF Replay yields an almost null EO by the end of the task stream.
\begin{table*}[htbp]
\centering
\caption{Results of the CL strategies on both datasets (CXP and NIH), considering both classification performance and fairness metrics. In \textbf{bold} is highlighted the best method for each metric in each dataset.}
\begin{tabular}{|c|c|c|c|c|c|c|}
\hline
\multicolumn{1}{|c|}{Dataset} & \multicolumn{3}{c|}{\textbf{CXP Dataset}} & \multicolumn{3}{c|}{\textbf{NIH Dataset}} \\ \hline
\textbf{Strategy\textbackslash{}Metric} & \textbf{AUC} & \textbf{Gender EO} & \textbf{Age EO} & \textbf{AUC} & \textbf{Gender EO} & \textbf{Age EO} \\ \hline
Joint Training & 0.78 & 0.008 & 0.148 & 0.78 & -0.010 & -0.024 \\ \hline
Fine-Tuning & 0.55 & 0.002 & 0.014 & 0.57 & 0.016 & -0.115 \\ \hline \hline
Replay & 0.60 & -0.013 & 0.065 & 0.60 & -0.022 & -0.005 \\ \hline
LwF & 0.68 & 0.028 & 0.059 & 0.65 & 0.013 & 0.046 \\ \hline
Pseudo-Label & \textbf{0.69} & \textbf{0.001} & 0.061 & \textbf{0.68} & \textbf{0.003} & 0.043 \\ \hline
Replay LwF & 0.67 & -0.002 & \textbf{0.023} & 0.65 & -0.021 & \textbf{-0.002} \\ \hline
\end{tabular}
\label{tab:results}
\end{table*}

\begin{figure}[thbp]
  \centering
    \includegraphics[width=0.35\textwidth]{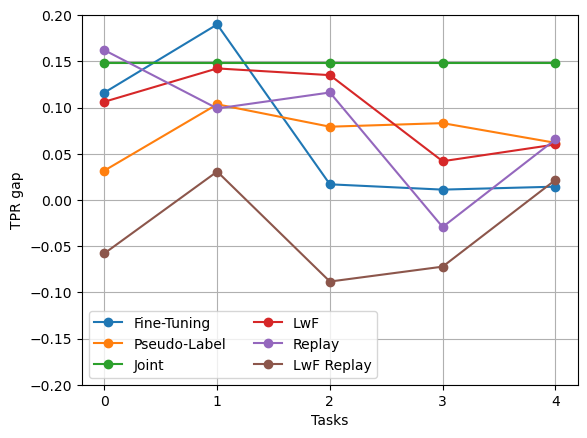}
    \caption{Age EO on CXP of all the considered CL strategies.}
    \label{fig:ageTPR-CXP}
\end{figure}

\begin{figure}[thbp]
  \centering
  \begin{subfigure}[b]{0.35\textwidth}
    \centering
    \includegraphics[width=\textwidth]{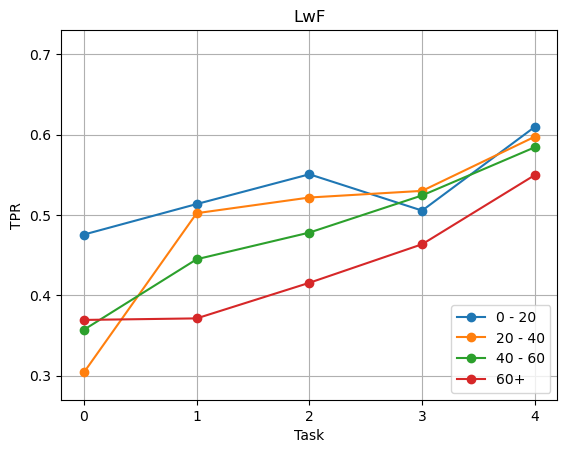}
    \caption{TPR of each age group considering the LwF approach.}
    \label{fig:LwF-ageTPR-CXP}
  \end{subfigure}
  \hfill
  \begin{subfigure}[b]{0.35\textwidth}
    \centering
    \includegraphics[width=\textwidth]{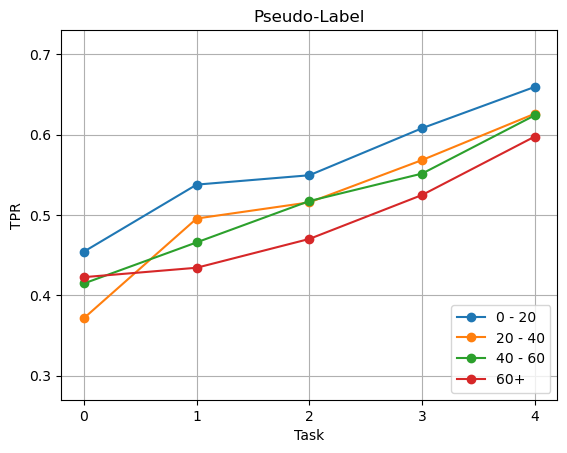}
    \caption{TPR of each age group considering the Pseudo-Label approach.}
    \label{fig:Pseudo-Label-ageTPR-CXP}
  \end{subfigure}
  \hfill
  \begin{subfigure}[b]{0.35\textwidth}
    \centering
    \includegraphics[width=\textwidth]{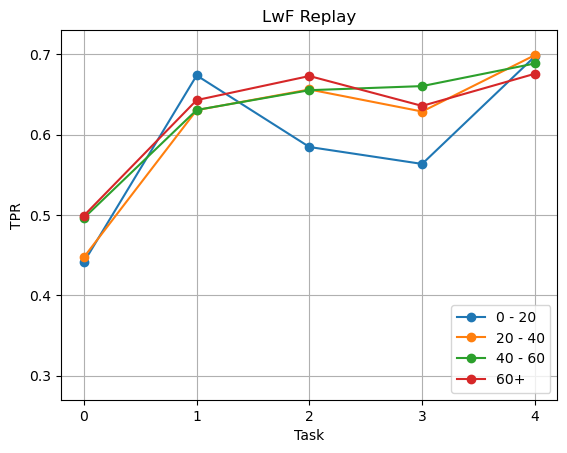}
    \caption{TPR of each age group considering the LwF Replay approach.}
    \label{fig:LwFReplay-ageTPR-CXP}
  \end{subfigure}
  
  \caption{TPR evolution relative to each age group, of
    the three best CL strategies.}
  \label{fig:results_ageTPR}
\end{figure}

\subsection{Analysis on the Fairness evolution on the age attribute}
\label{subsec:results_fairness_age}
Lastly, we analyze the performance of the different strategies across the age groups defined in Sec. \ref{subsec:evaluation_metrics}. On the CXP dataset, joint training shows the highest TPR for the 0–20 group and the lowest for the 60+ group, with a gap of 0.15. Fig. \ref{fig:ageTPR-CXP} plots this gap across the task stream for all strategies. TPR results for all age groups using the top three methods are shown in Fig. \ref{fig:results_ageTPR}.

From the plots we can notice that, considering LwF and Pseudo-Label, after training on all tasks, the TPR is the highest on people younger than 20 and the lowest on people older than 60. Moreover, the two methods display very similar EOs: the difference between the highest TPR and the lowest is around 0.06 for both LwF and Pseudo-Label. When considering the LwF Replay approach, we observe that the final gap is very small, taking the value of 0.023.

\subsection{Overall considerations}
In Table \ref{tab:results} the results of all strategies on both datasets are reported. Overall, the LwF Replay approach is the best in terms of age gap; however, its suboptimality in terms of classification performance, on both datasets, and the gender EO on the NIH dataset limit its employability. On the other hand, Pseudo-Label performs better in terms of AUC and gender EO, exhibiting a slightly higher value in terms of age EO. In other words, Pseudo-Label exhibits the best combination of results. 

It's worth mentioning that, in the case of the LwF Replay approach, the most favored and disfavored age groups do not correspond to the age EO results of the models resulted from the joint training. Moreover, the gender EO gap favoring males observed in the results of the LwF strategy is not present in the static setting in which the joint model was trained. This further emphasizes the unpredictability of fairness results when considering a CL scenario, hence the need of considering fairness metrics in these settings.

\section{Conclusion and Future Work} \label{sec:ConclusionFutureWork}

In this study, we leveraged continual learning (CL) techniques to address the medical image diagnosis problem. Specifically, we explored a class-incremental learning (CIL) scenario where new diseases are introduced incrementally and assessed how biases evolve as the model adapts. We observed that traditional approaches like Replay struggle to retain past knowledge, whereas LwF and Pseudo-Label outperform Replay and LwF Replay, with Pseudo-Label showing slightly better overall performance.

We further evaluated the fairness of CL methods by analyzing Equality of Opportunity (EO) between male and female groups, and among different age groups. Results show that Pseudo-Label exhibits the best EO regarding gender, and achieves the highest classification performance while maintaining reasonable fairness across age groups. Conversely, LwF and LwF Replay exhibit greater gender bias and slightly lower AUC values. Thus, Pseudo-Label emerges as a promising CL method for medical image diagnosis, balancing classification performance and fairness.

While our findings are significant, further research is needed to fully understand and mitigate biases in CL applications. Although LwF and Pseudo-Label help reduce forgetting, a considerable performance gap remains compared to static training. Evaluating additional methods will be crucial to improving overall performance while preserving fairness. Moreover, while our analysis focuses on a CIL setting, real-world medical applications may involve more complex scenarios worth exploring.
As part of future work, we also plan to integrate and systematically evaluate fairness-aware techniques within this continual learning setting, aiming to close the observed gap in performance across demographic groups.
Overall, this study serves as a foundational exploration, encouraging further investigation into diverse and intricate CL scenarios to establish robust benchmarks for fairness evolution analysis.

\bibliography{ifacconf} 
\appendix
\input{appendix}

\end{document}

%% file: appendix.tex
\section{Label distribution for CXP and NIH dataset}
\label{app:labelDistribution}
We provide a visual representation of the frequency of each pathology across tasks for the CXP and NIH datasets, respectively. 
The blue bars correspond to the pathologies associated with the current task, while the light blue bars correspond to the other pathologies, and the blue contour represents the frequency of each disease in the original dataset. During each task, we keep all the images in the dataset containing at least one of the pathologies associated to the task; however, other diseases may be present even though the information on the presence of such pathologies is not available, hence they’re hidden pathologies.

\begin{figure*}[thbp]
  \centering
    \centering
    \includegraphics[width=0.6\textwidth]{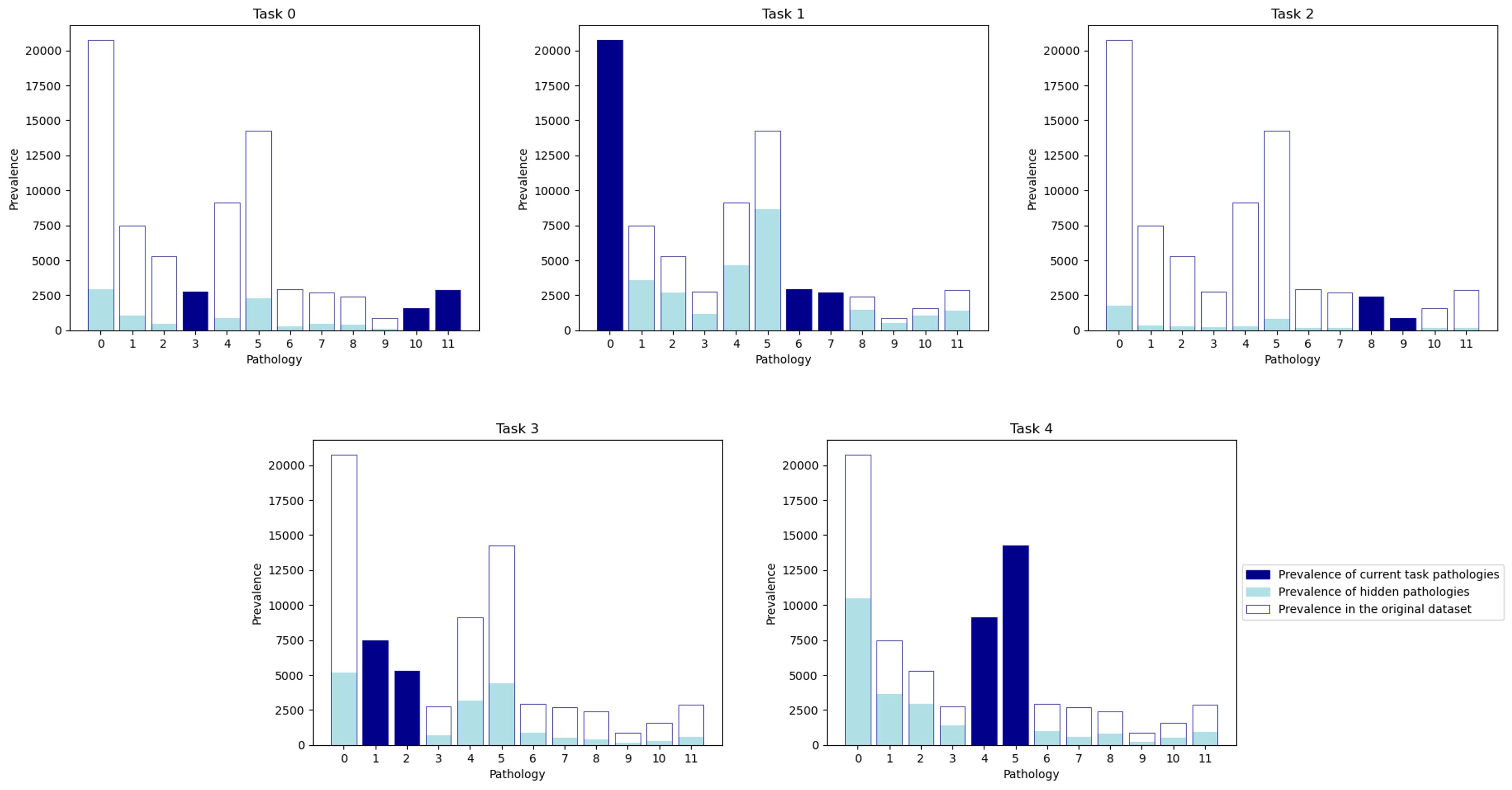}
    \label{fig:componentCounts1}

  \caption{Visual representation of the frequency of each pathology in each task on CXP.}
  \label{fig:results_TPR}
\end{figure*}

\begin{figure*}[!h]
  \centering
    \centering
    \includegraphics[width=0.6\textwidth]{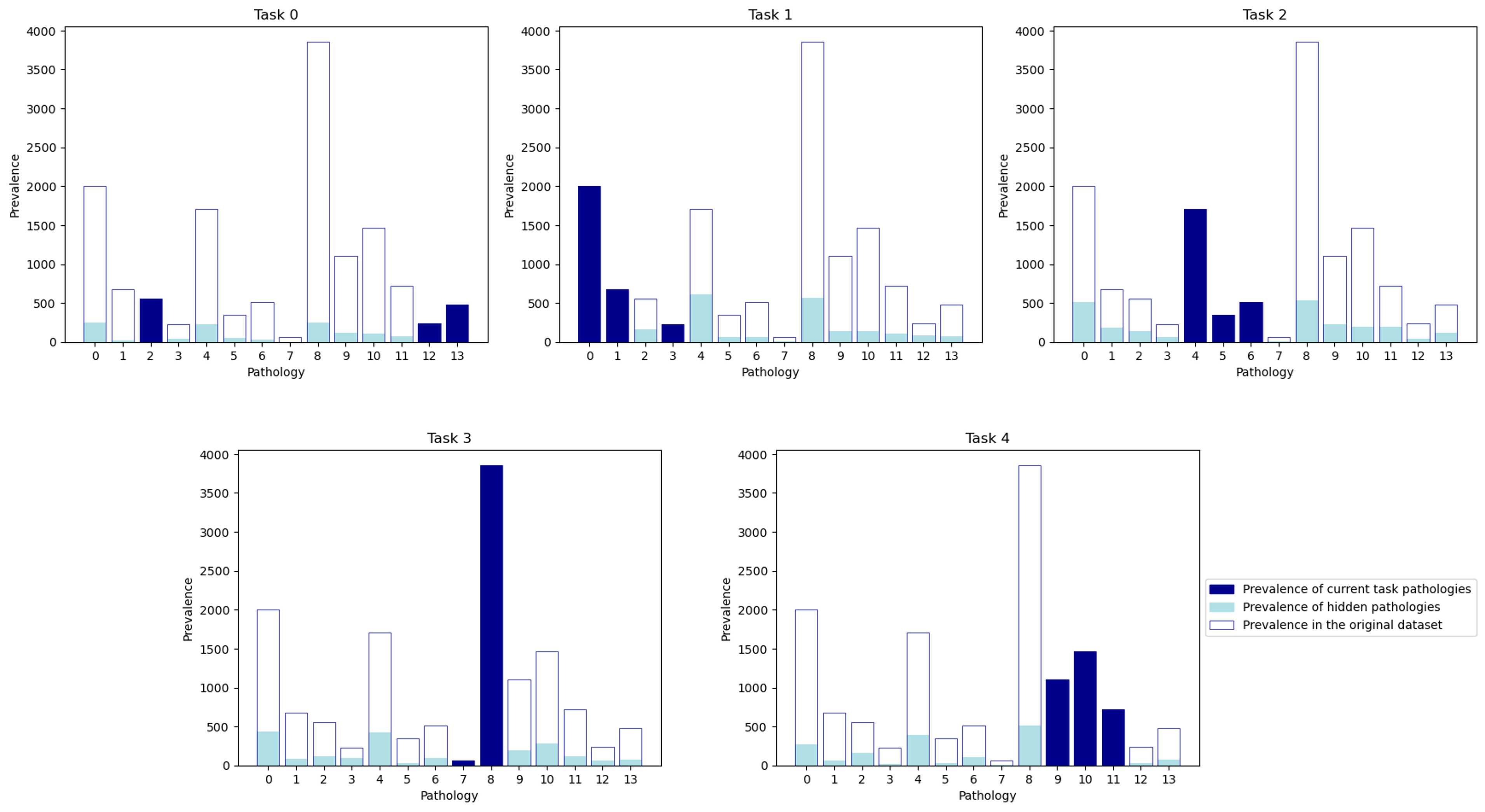}
    \label{fig:componentCountsNIH}
    \caption{Visual representation of the frequency of each pathology in each task on NIH.}
\end{figure*}

In the case of CXP, Task 0 contains information on the classes Consolidation, Pneumonia, and Pneumothorax, Task 1 involves Lung Opacity, Enlarged Cardiomediastinum, and Fracture, Task 2 considers Lung Lesion and Pleural Other, while Task 3 includes Atelectasis and Cardiomegaly and finally Task 4 revolves around Edema and Effusion.

Instead, concerning the NIH dataset, Task 0 contains information on the classes Consolidation, Pneumonia, and Pneumothorax, Task 1 involves Atelectasis, Cardiomegaly, Edema, Task 2 considers Effusion, Emphysema and Fibrosis, while Task 3 includes Hernia and Infiltration and finally Task 4 revolves around Mass, Nodule and Pleural Thickening.

\section{Gender frequency for each task}
\label{app:genderFrequency}
\begin{figure*}[thbp]
  \centering

  \begin{subfigure}[b]{0.25\textwidth}
    \centering
    \includegraphics[width=\textwidth]{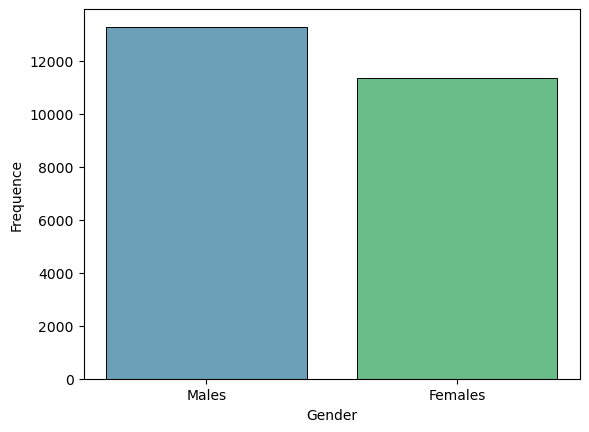}
  \end{subfigure}
  
  \begin{subfigure}[b]{0.6\textwidth}
    \centering
    \includegraphics[width=\textwidth]{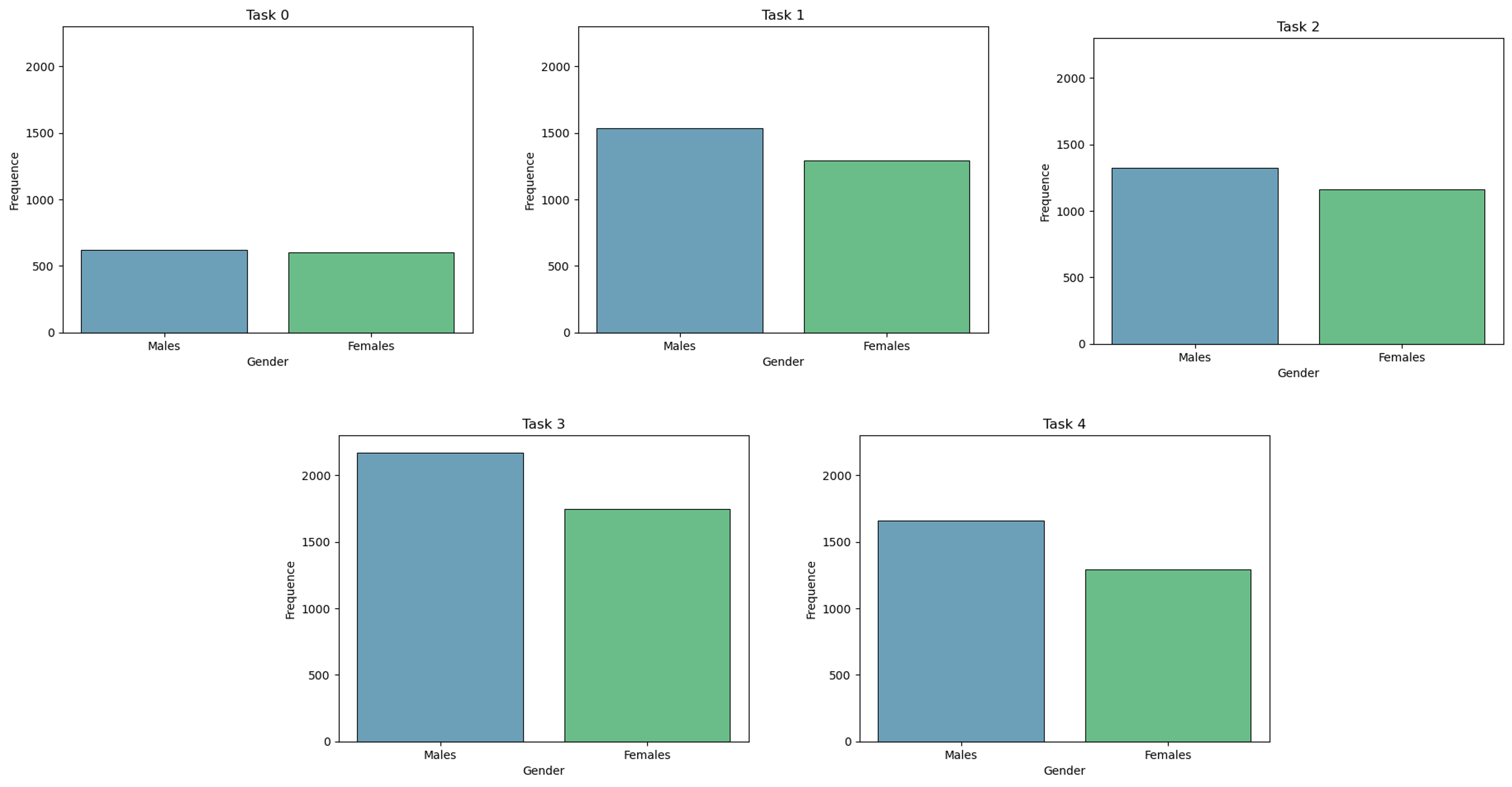}
  \end{subfigure}
  
  \caption{Visual representation of the frequency of the two genders in the whole dataset (on the top) and in all tasks (on the bottom), considering the CXP dataset.}
  \label{fig:femaleCounts}
\end{figure*}
\begin{figure*}[thbp]
  \centering

  \begin{subfigure}[b]{0.25\textwidth}
    \centering
    \includegraphics[width=\textwidth]{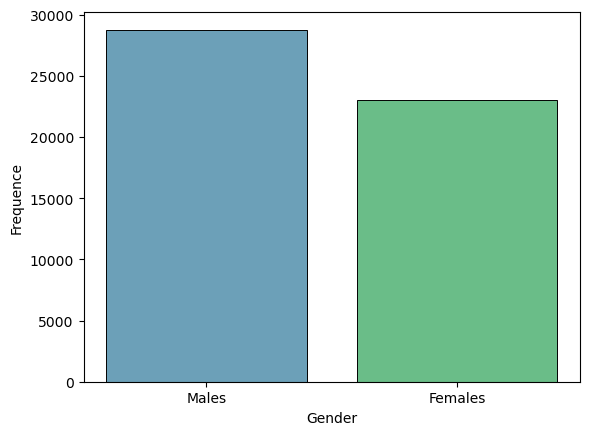}
  \end{subfigure}
  
  \begin{subfigure}[b]{0.6\textwidth}
    \centering
    \includegraphics[width=\textwidth]{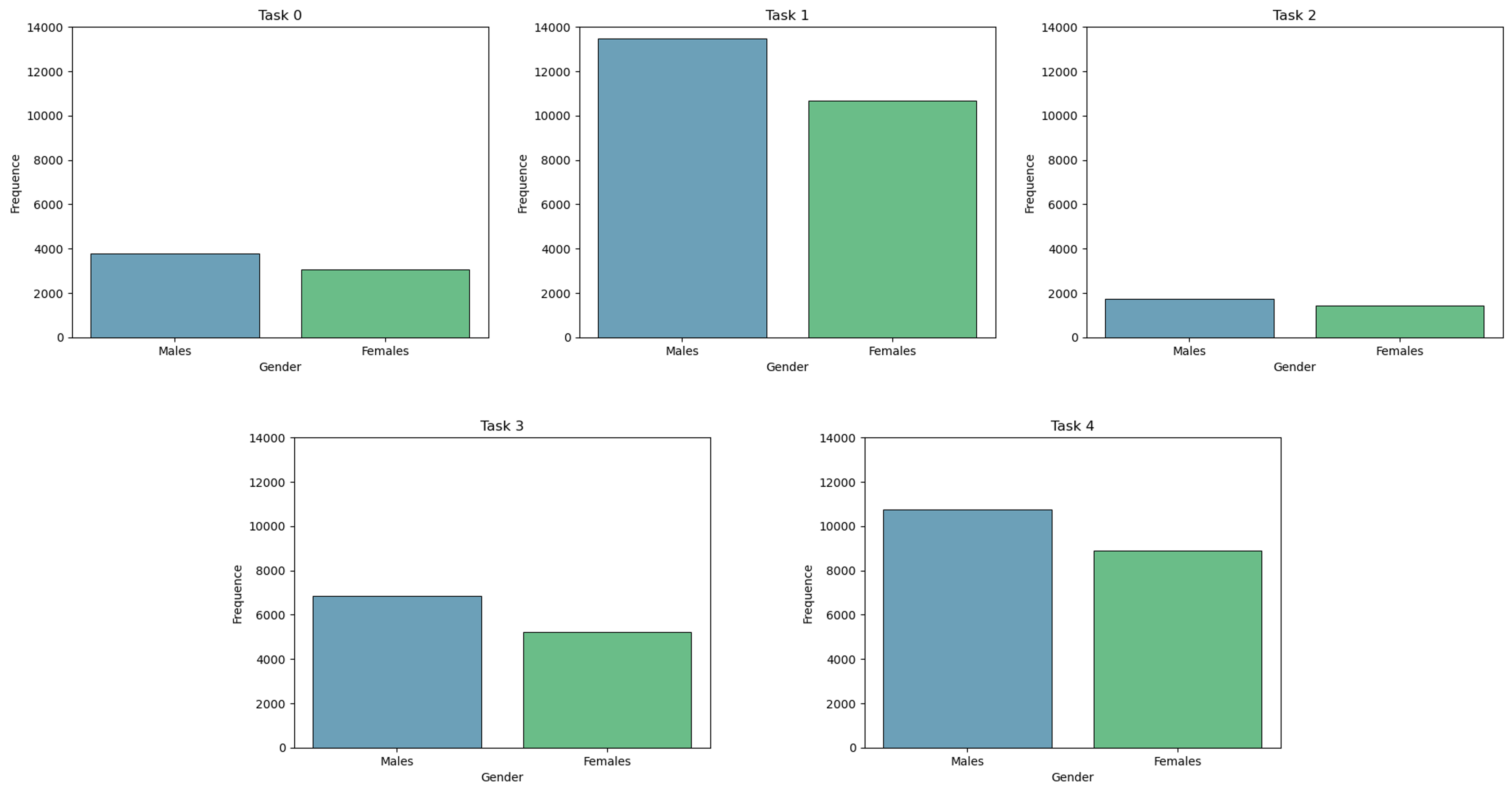}
  \end{subfigure}
  
  \caption{Visual representation of the frequency of the two genders in the whole dataset (on the top) and in all tasks (on the bottom), considering the NIH dataset.}
  \label{fig:femaleCountsNIH}
\end{figure*}

We present a visual representation of gender distribution across the datasets and individual tasks. The top section of Fig. \ref{fig:femaleCounts} depicts the overall frequency of the two genders in the entire CXP dataset, while the bottom section shows their distribution across all tasks. Similarly, Fig. \ref{fig:femaleCountsNIH} provides the corresponding visualization for the NIH dataset.

\section{Age groups frequency for each task}
\label{app:ageFrequency}
\begin{figure*}[thbp]
  \centering

  \begin{subfigure}[b]{0.25\textwidth}
    \centering
    \includegraphics[width=\textwidth]{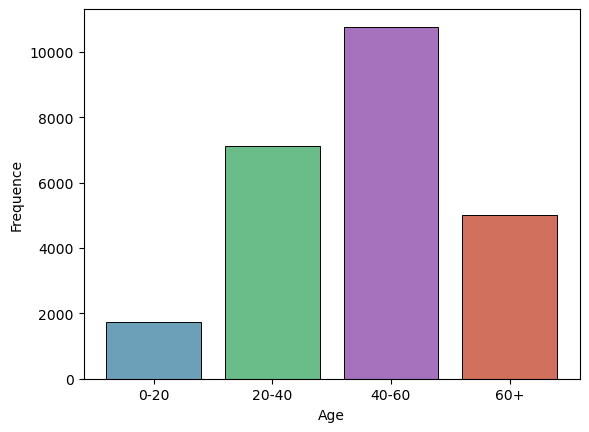}
  \end{subfigure}
  
  \begin{subfigure}[b]{0.6\textwidth}
    \centering
    \includegraphics[width=\textwidth]{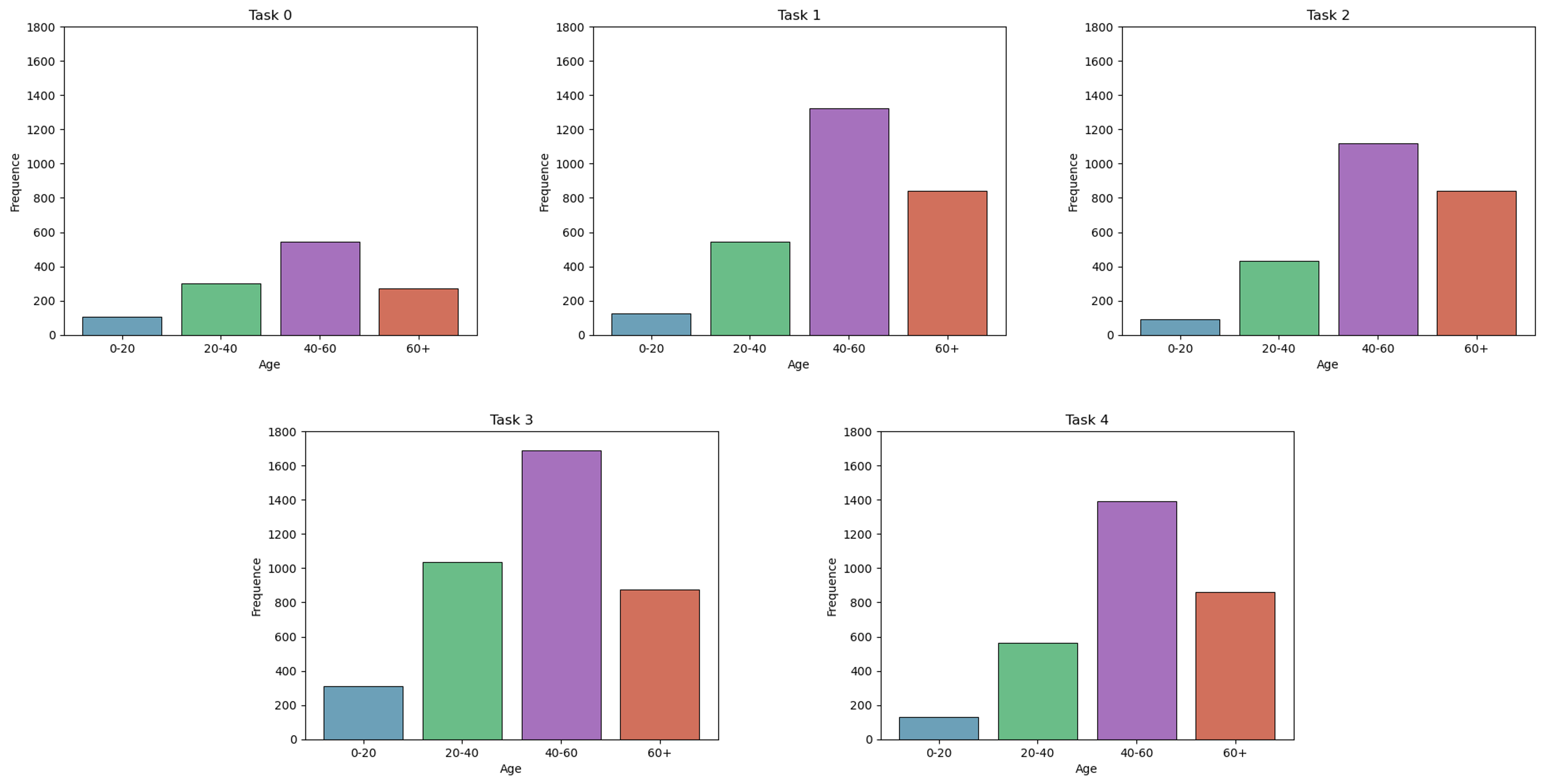}
  \end{subfigure}
  
  \caption{Visual representation of the frequency of the four age groups in the whole dataset (on the top) and in all tasks (on the bottom), considering the CXP dataset.}
  \label{fig:ageCounts}
\end{figure*}

\begin{figure*}[thbp]
  \centering

  \begin{subfigure}[b]{0.25\textwidth}
    \centering
    \includegraphics[width=\textwidth]{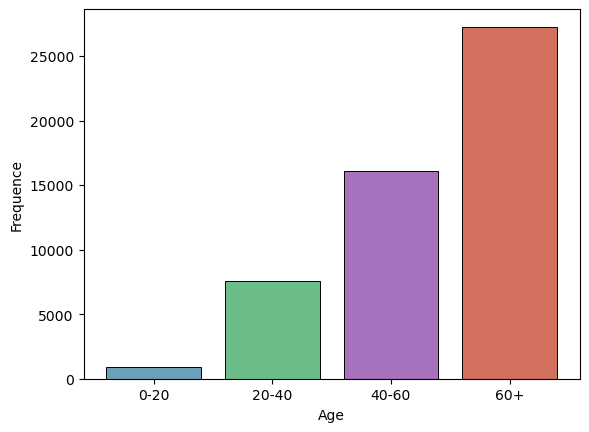}
  \end{subfigure}
  
  \begin{subfigure}[b]{0.6\textwidth}
    \centering
    \includegraphics[width=\textwidth]{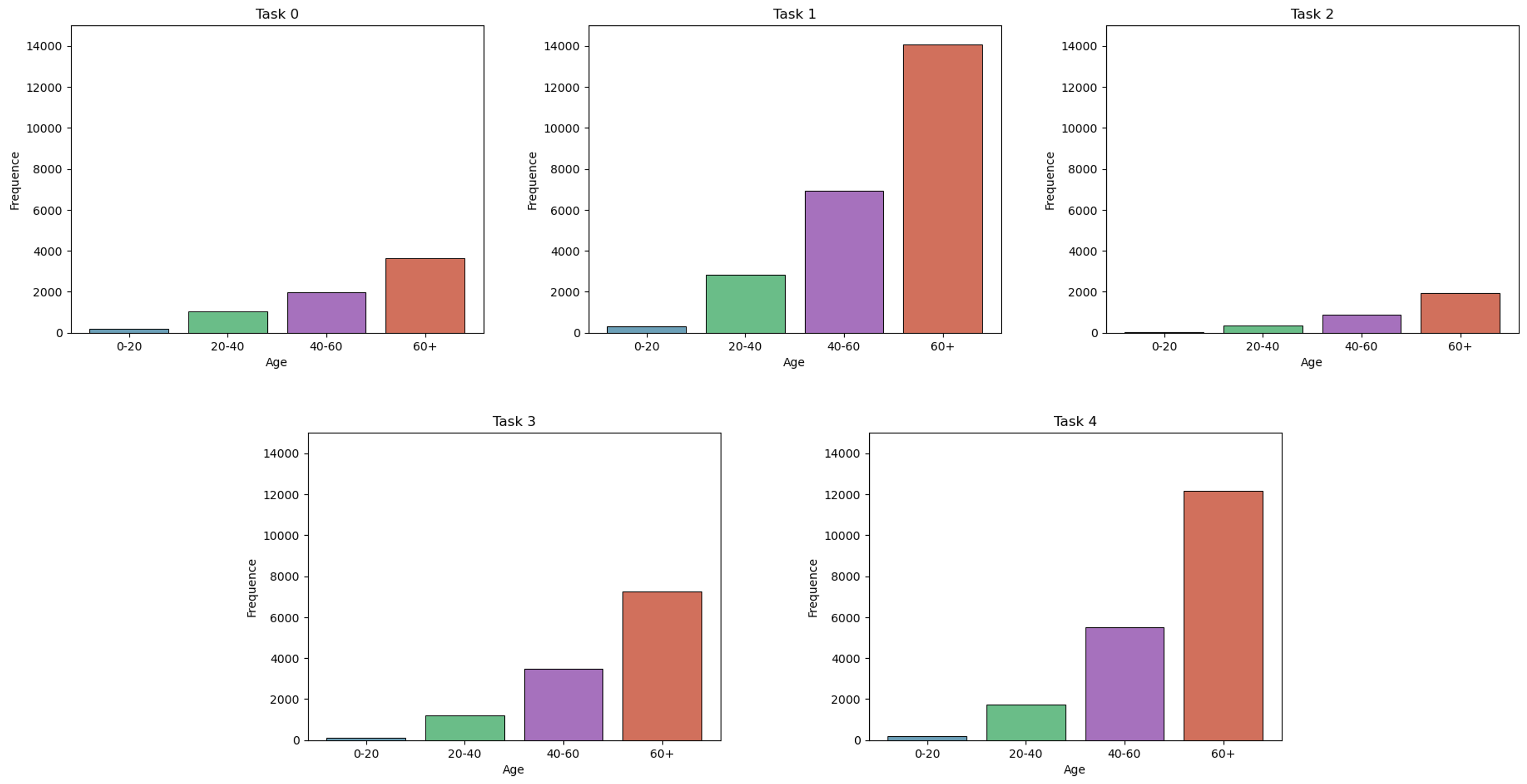}
  \end{subfigure}
  
  \caption{Visual representation of the frequency of the four age groups in the whole dataset (on the top) and in all tasks (on the bottom), considering the NIH dataset.}
  \label{fig:ageCountsNIH}
\end{figure*}

We present a visual representation of the frequency distribution of four age groups across the datasets. The top section of Fig. \ref{fig:ageCounts} depicts the overall frequency of the four groups in the entire CXP dataset, while the bottom section shows their distribution across all tasks. Similarly, Fig. \ref{fig:ageCountsNIH} provides the corresponding visualization for the NIH dataset.

\section{Pseudocode of LwF and Pseudo-Label}
\label{app:pseudocodes}
\subsection{Learning without Forgetting}
Learning without Forgetting (LwF) is a distillation-based technique designed to transfer knowledge from previous tasks to new ones. During training on a new task, the model is not only optimized to make accurate predictions on the current data but also to replicate its own predictions from earlier tasks. This dual objective helps mitigate catastrophic forgetting, enabling the model to retain previously learned knowledge while adapting to new information.

\begin{algorithm}
\caption{Learning without forgetting}
\begin{algorithmic}[1]
    \Require Current task dataset $D_{\text{new}}$, previous model parameters $\theta_{\text{old}}$
    \Ensure Updated model parameters $\theta_{\text{new}}$ for the current task
    \State Initialize model parameters $\theta_{\text{new}}$
    \For{$(X_{\text{new}}, Y_{\text{new}})$ \textbf{in} $D_{\text{new}}$}
        \State $Y_{\text{old}} \gets f_{\theta_{\text{old}}}(X_{\text{new}})$
        \State $\hat{Y}_{\text{old}}, \hat{Y}_{\text{new}} \gets f_{\theta_{\text{new}}}(X_{\text{new}})$
        \State $L = L(Y_{\text{new}}, \hat{Y}_{\text{new}}) + \lambda \cdot L_{\text{KD}}(Y_{\text{old}}, \hat{Y}_{\text{old}})$ 
        \State Update model parameters: $\theta_{\text{new}} \gets \theta_{\text{new}} - \eta \cdot \nabla_{\theta_{\text{new}}} L$  

    \EndFor
    \State \textbf{return} $\theta_{\text{new}}$
\end{algorithmic}
\end{algorithm}

\subsection{Pseudo-Label}
Similar to LwF, Pseudo-Label leverages a model trained on past tasks to transfer knowledge to the model being trained on a new task. This approach is particularly useful in multilabel settings, where classes from previous tasks may still be present in new task samples, but their corresponding labels are unavailable.

To address this, the previously trained model is used to infer the presence of old classes in new task samples. For each old class, the model outputs a probability indicating its presence in the current sample. These probabilities are then binarized using a confidence threshold $\tau$, and the corresponding ground-truth targets are updated accordingly.

Finally, the model is trained on the input samples using the modified ground-truth targets, allowing it to incorporate past knowledge while learning new tasks. 

\begin{algorithm}
\caption{Pseudo-Label Algorithm}
\begin{algorithmic}[1]
    \Require Current task dataset $D_{\text{new}}$, previous model parameters $\theta_{\text{old}}$, set of old classes $L_{old}$, threshold $\tau$
    \Ensure Updated model parameters $\theta_{\text{new}}$ for the current task
    \State Initialize model parameters $\theta_{\text{new}}$
    \For{$(X_{\text{new}}, Y_{\text{new}})$ \textbf{in} $D_{\text{new}}$}
        \State $\hat{Y}_{\text{old}} \gets f_{\theta_{\text{old}}}(X_{\text{new}})$
        \State $ Y_{\text{old}} \gets \emptyset$
        \For{ $ (x_{\text{new}}, \hat{y}_{\text{old}})$ in $(X_{\text{new}}, \hat{Y}_{\text{old}} )   $}
            \State $y_{\text{old}} \gets \emptyset$
            \For{$~l$ in $L_{old}$}
                \If{$\hat{y}_{\text{old}}^{l} \geq \tau$}
                    \State $y_{\text{old}}^{l} \gets 1$
                \Else
                    \State $y_{\text{old}}^{l} \gets 0$
                \EndIf
            \EndFor
            \State $ Y_{\text{old}} \gets Y_{\text{old}} \cup y_{\text{old}} $
        \EndFor
        \State $ Y \gets Y_{\text{old}} \cup Y_{\text{new}} $
                
        \State $\hat{Y} \gets f_{\theta_{\text{new}}}(X_{\text{new}})$ 

        \State $L = L(Y, \hat{Y}) $

        \State $\theta_{\text{new}} \gets \theta_{\text{new}} - \eta \cdot \nabla_{\theta_{\text{new}}} L$  

    \EndFor
    \State \textbf{return} $\theta_{\text{new}}$
\end{algorithmic}
\end{algorithm}

\section{Results on NIH}
\begin{figure}[thbp]
  \centering

    \centering
    \includegraphics[width=0.35\textwidth]{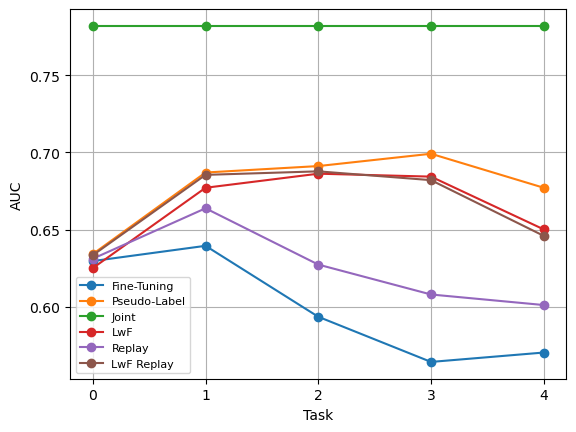}
    \caption{AUC metric, evaluated on each strategy, averaged on all the pathologies seen so far (NIH).}
    \label{fig:AUC_ROC_NIH}
\end{figure}

As is notable from Fig. \ref{fig:AUC_ROC_NIH}, the average AUC on the model trained in the joint training on NIH is 0.78.

As it was observed in the case of the CXP dataset, the Fine-Tuning approach fails at maintaining the knowledge of previous tasks. Indeed, the trend on the overall average AUC is strongly decreasing. Similarly, Replay performs poorly in this scenario, exhibiting only a small improvement with respect to the Fine-Tuning approach. 

In particular, Replay achieves a final ROC AUC of 0.60, compared to the 0.57 achieved by the Fine-Tuning approach. 

On the other hand, the LwF, Pseudo-Label and LwF Replay strategies perform well on this dataset. The main difference with respect to the results on the CXP dataset is that LwF Replay performs very similarly to LwF, and both exhibit a slightly lower AUC with respect to Pseudo-Label. Indeed, the final value of AUC of both LwF Replay and LwF is 0.65, while the final value of the Pseudo-Label strategy is 0.68. As it was for the CXP dataset, there is a notable gap between the optimal performance represented by the joint training strategy, which achieves a final value of 0.78, and the optimal CL strategy, i.e., Pseudo-Label.

\begin{figure}[htbp]
  \centering
  \begin{subfigure}[b]{0.35\textwidth}
    \includegraphics[width=\textwidth]{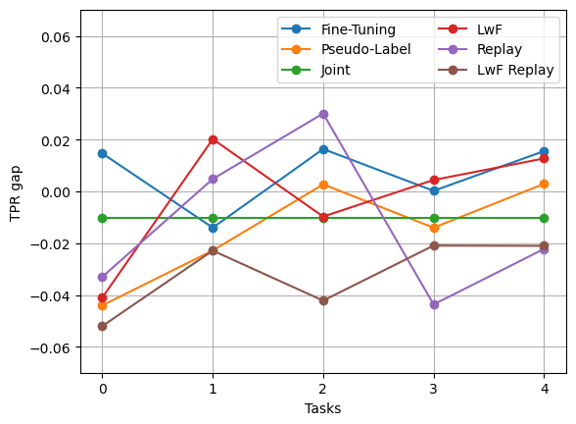}
    \caption{Gender EO on NIH of all the considered CL strategies.}
    \label{fig:TPR_NIH}
  \end{subfigure}
  \hfill
  \begin{subfigure}[b]{0.35\textwidth}
    \includegraphics[width=\textwidth]{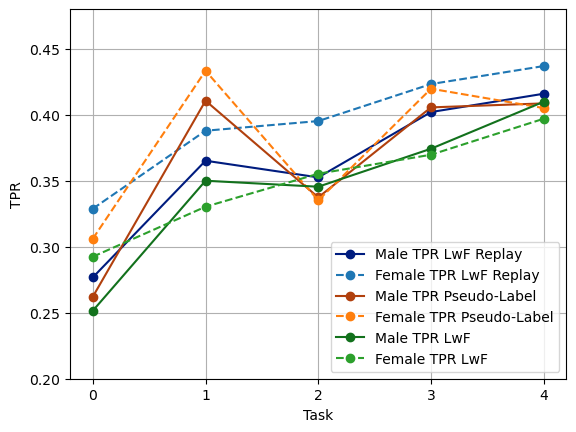}
    \caption{Male and female TPR of the three best CL strategies.}
    \label{fig:sex_TPR-NIH}
  \end{subfigure}
  \caption{Fairness metric results on NIH.}
  \label{fig:summaryTPR_NIH}
\end{figure}

Concerning the gender EO, as mentioned in the previous sections, previous works had found that the models trained on NIH were biased toward males (\cite{seyyedkalantari2020chexclusionfairnessgapsdeep}). In the case of NIH, we find that the TPR is slighlty higher for females, and it takes the value of $-0.010$, where the minus indicates that females are the advantaged group.

As previously stated, there are many factors that may contribute to the difference in results with respect to the SOTA, for example the choice of only keeping one image per patient and of not considering ``No Finding'' images. 

In Fig. \ref{fig:TPR_NIH}, we display the plots of the gap disparity between males and females for all strategies. As done in the case of the CXP dataset, we focus on the TPR of the three best methods: LwF, Pseudo-Label and LwF Replay. The plots relative to the male and female TPR of these approaches are reported in Fig. \ref{fig:sex_TPR-NIH}.

From the figure we can notice that, as it was for the CXP dataset, the model resulting from training on all tasks using the Pseudo-Label strategy displays an almost null EO, while instead the LwF approach slightly favours the performance on males. However, the EO of the LwF approach is smaller with respect to the one observed on the CXP dataset. On the other hand, the use of the LwF Replay approach results in an EO favoring the performance on females.

\begin{figure}[thbp]
  \centering
    \includegraphics[width=0.35\textwidth]{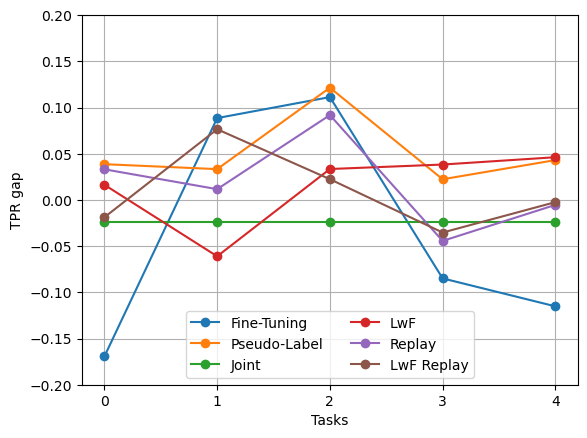}
    \caption{Age EO on NIH of all the considered CL strategies.}
    \label{fig:ageTPR-NIH}
\end{figure}

\begin{figure}[thbp]
  \centering
  
  \begin{subfigure}[b]{0.35\textwidth}
    \centering
    \includegraphics[width=\textwidth]{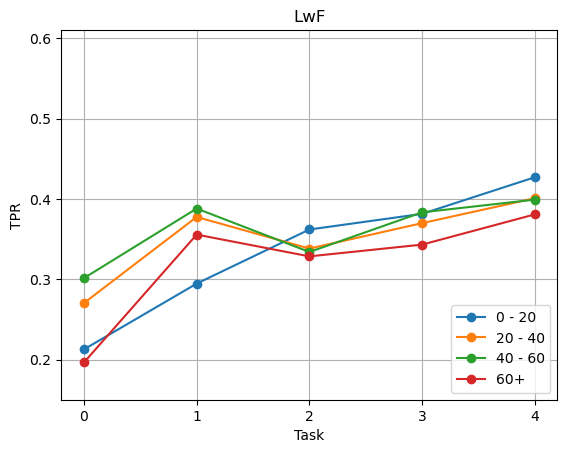}
    \caption{TPR of each age group considering the LwF approach.}
    \label{fig:LwF-ageTPR-NIH}
  \end{subfigure}
  \hfill
  \begin{subfigure}[b]{0.35\textwidth}
    \centering
    \includegraphics[width=\textwidth]{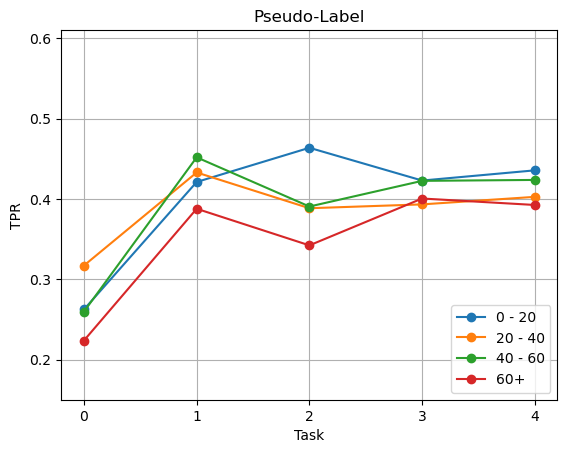}
    \caption{TPR of each age group considering the Pseudo-Label approach.}
    \label{fig:Pseudo-Label-ageTPR-NIH}
  \end{subfigure}
  \hfill
  \begin{subfigure}[b]{0.35\textwidth}
    \centering
    \includegraphics[width=\textwidth]{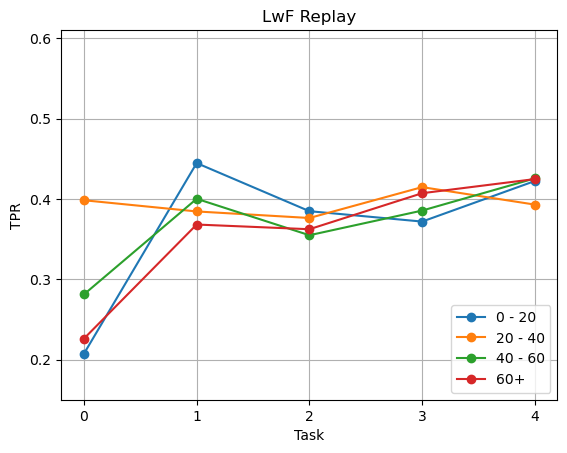}
    \caption{TPR of each age group considering the LwF Replay approach.}
    \label{fig:LwFReplay-ageTPR-NIH}
  \end{subfigure}
  
  \caption{TPR evolution relative to each age group, of
    the three best CL strategies.}
  \label{fig:results_ageTPR_NIH}
\end{figure}

Additionally, we report the results of TPR and EO relative to each age group.
The results on the joint training show that the group with the highest TPR is patients between 40 and 60, while the most unfavored group is patients younger than 20, and the gap is of 0.053.

As depicted in the previous sections, we define the EO as the difference between the TPR of the youngest and the oldest group, and we plot it in Fig. \ref{fig:ageTPR-NIH}, for each strategy, after training on each task.
Instead, the results on the TPR for all age groups relative to the three best methods (LwF, Pseudo-Label and LwF Replay) are displayed in Fig. \ref{fig:results_ageTPR_NIH}.

From the plots we can notice that, considering the LwF and Pseudo-Label approaches, after training on all tasks, the TPR is the highest on people younger than 20 and the lowest on people older than 60. Moreover, the two strategies display very similar gaps: in the case of LwF the gap is 0.046, while considering Pseudo-Label it's 0.043. The two gaps are slightly smaller with respect to the ones noticed in the CXP dataset.

When considering the LwF Replay approach, we observe that the disparity between the most and the least advataged groups is marginally smaller: it takes the value of 0.032, favoring patients older than 60 and disfavoring patients between 20 and 40. On the other hand, the TPRs on the youngest and oldest groups are very similar, hence the EO in this case is $-0.002$.